\documentclass[aps,prl,groupedaddress, letterpaper, twocolumn, showpacs]{revtex4}
\usepackage{amsmath}
\usepackage{amssymb}
\usepackage{graphicx}
\usepackage{color}
\usepackage{tikz}
\usepackage{pgffor}
\usepackage{verbatim}
\usepackage{pstricks}
\usepackage{pst-3dplot}
\newcommand{\be}{\begin{equation} }
\newcommand{\ee}{\end{equation} }
\newcommand{\ba}{\begin{eqnarray} }
\newcommand{\ea}{\end{eqnarray} }

\newcommand{\bpm}{\begin{pmatrix}}
\newcommand{\epm}{\end{pmatrix}}
\newcommand{\bmm}{\begin{matrix}}
\newcommand{\emm}{\end{matrix}}

\begin{document}

\title{Braiding statistics of loop excitations in three dimensions}

\author{Chenjie Wang}

\author{Michael Levin}

\affiliation{James Franck Institute and Department of Physics, University of Chicago, Chicago, Illinois 60637, USA}

\date{\today}

\begin{abstract}
While it is well known that three dimensional quantum many-body systems can support non-trivial braiding statistics between
particle-like and loop-like excitations, or between two loop-like excitations, we argue that a more fundamental quantity is the
statistical phase associated with braiding one loop $\alpha$ around another loop
$\beta$, while both are linked to a third loop $\gamma$. We study this three-loop braiding in the context of
$(\mathbb{Z}_N)^K$ gauge theories which are obtained by gauging a gapped, short-range entangled lattice boson model
with $(\mathbb{Z}_N)^K$ symmetry. We find that different short-range entangled bosonic states with the same
$(\mathbb{Z}_N)^K$ symmetry (i.e. different symmetry-protected topological phases) can be distinguished by their
three-loop braiding statistics.
\end{abstract}


\maketitle

\begin{figure}[b]
\centering
\includegraphics{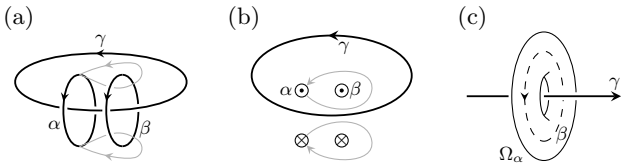}
\caption{(a) Three-loop braiding process. The gray curves show the paths of two points on the moving loop $\alpha$.
(b) A top view of the braiding process within the plane that $\gamma$ lies in. (c) A torus $\Omega_\alpha$ is swept out by $\alpha$ during
the braiding. Loop $\beta$ (dashed circle) is enclosed by $\Omega_\alpha$.}
\label{fig_braiding}
\end{figure}

{\it Introduction.}
A powerful way to characterize the topological properties of two dimensional gapped quantum many-body systems is to examine their quasiparticle braiding statistics \cite{anyon}. It is thus natural to wonder: what is the analogous quantity that characterizes three dimensional (3D) systems? The simplest candidate --- 3D quasiparticle statistics --- is of limited use since 3D systems can only support bosonic and fermionic quasiparticles. On the other hand, 3D systems can support much richer braiding statistics between particle-like excitations and loop-like excitations\cite{aharonov59, alford89, krauss89} or between two loop-like excitations\cite{aneziris91, alford92, baez07}. Thus, one might guess that particle-loop and loop-loop braiding statistics are the natural generalizations of quasiparticle statistics to three dimensions.


In this paper, we argue that this guess is incorrect: particle-loop and loop-loop braiding statistics do not fully
capture the topological structure of 3D many-body systems. Instead, more complete information can be obtained by
considering a \emph{three-loop} braiding process in which a loop $\alpha$ is braided around another loop $\beta$, while
both are linked with a third loop $\gamma$ (Fig.~\ref{fig_braiding}). We believe that three-loop braiding statistics is
one of the basic pieces of topological data that describe 3D gapped many-body systems, and much of this work is devoted to understanding the
general properties of this quantity. Also, as an application, we show that three-loop statistics can be used to distinguish different short-range
entangled many-body states with the same (unitary) symmetry --- i.e. different {\it symmetry-protected
topological (SPT) phases}\cite{chen13, 1dspt, hasan10}. The latter result shows that the braiding statistics approach to SPT phases, outlined in
Ref.~\onlinecite{levin12}, can be extended to three dimensions.

{\it Discrete gauge theories.}
For concreteness, we focus our analysis on a simple 3D system with loop-like excitations,
namely lattice $(\mathbb{Z}_N)^K$ gauge theory\cite{footnote0}.
More specifically, we consider a 3D lattice boson model built out of $K$ different species of bosons, where the number of bosons in each species is
conserved \emph{modulo} $N$ so that the system has a $(\mathbb Z_{N})^K$ symmetry. We suppose that the ground state
of the boson model is gapped and short-range entangled --- that is, it can be transformed into a product state
by a local unitary transformation\cite{LUT}. We then imagine coupling such a lattice boson model to a
$(\mathbb Z_{N})^K$ lattice gauge field\cite{kogut79}.

In general these gauge theories contain two types of excitations: point-like ``charge'' excitations which carry gauge charge,
and string-like ``vortex loop'' excitations which carry gauge flux. The most general charge excitations can carry gauge charge
$q = (q_1,...,q_K)$ where each component $q_m$ is an integer defined modulo $N$. The most general vortex loop can carry gauge flux
$\phi = (\phi_1,...,\phi_K)$ where $\phi_m$ is a multiple of $\frac{2\pi}{N}$. In fact, since we
can always attach a charge to a vortex loop to obtain another vortex loop, a general vortex loop excitation
carries {\it both} flux and charge.

Let us try to understand the braiding statistics of these excitations. In general, there are three types of braiding processes
we can consider: processes involving two charges,
processes involving a charge and a loop, and processes involving multiple loops. Clearly, the first type of process cannot give any
statistical phase since the charges are excitations of the short-range entangled boson model and therefore must be bosons. On the
other hand, the second kind of process, involving charges and loops, \emph{can} give a
nontrivial phase. More specifically, if we braid a charge $q = (q_1,...,q_K)$ around a vortex loop with gauge flux
$\phi = (\phi_1,...,\phi_K)$, the resulting statistical phase is given by the Aharonov-Bohm formula
\begin{equation}
\theta = q \cdot \phi, \label{particlestat}
\end{equation}
where ``$\cdot$'' denotes the vector dot product.

\begin{figure}
\includegraphics[width=3.3in]{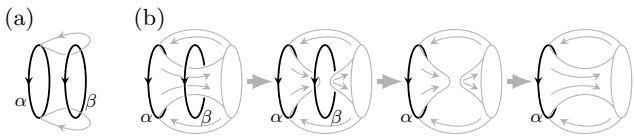}
\caption{(a) Braiding of two loops $\alpha, \beta$. (b) If $\alpha, \beta$ are neutral, the two-loop process can be smoothly deformed
into a process in which $\alpha$ is braided around the vacuum.}
\label{fig_twoloop}
\end{figure}

All that remains is to examine the braiding statistics of loops. The simplest process one can consider\cite{footnote1} involves braiding a loop $\alpha$
around another loop $\beta$ as shown in Fig.~\ref{fig_twoloop}(a).  To analyze this process, we use two facts about {\it unlinked} vortex loops: First,
a subset of vortex loops, which we call ``neutral'' loops, can be shrunk to a point and annihilated by local gauge invariant operators. Second, all other
vortex loops can be obtained from neutral loops by attaching an appropriate amount of charge.
With these facts in mind, let us first suppose that both $\alpha, \beta$ are neutral.
In this case, it follows from general principles that the
statistical phase $\theta_{\alpha\beta}=0$, since we can ``smoothly''\cite{footnote3} deform the two-loop braiding process into another process in which $\alpha$ is
braided around the vacuum [Fig.~\ref{fig_twoloop}(b)]. Now consider the general case where $\alpha, \beta$ carry charge. In this case, $\alpha, \beta$ can be thought of as
neutral loops with some attached charge. It then follows from the Aharonov-Bohm formula (\ref{particlestat}) that the Berry phase associated with braiding $\alpha$ around $\beta$ is
\begin{equation}
\theta_{\alpha \beta} = q_\alpha \cdot \phi_\beta + q_\beta \cdot \phi_\alpha, \label{twostringstat}
\end{equation}
where $q_\alpha, q_\beta$ and $\phi_\alpha, \phi_\beta$ denote the charge and flux carried by $\alpha, \beta$ respectively. To see this, note that during the two-loop braiding, the charge $q_\alpha$ is braided around the flux $\phi_\beta$ and the flux $\phi_\alpha$ is braided around the charge $q_\beta$.


While the above calculations show that 3D gauge theories can exhibit nonvanishing braiding statistics,
we can see that these statistical phases are the same for all gauge theories with gauge group $(\mathbb{Z}_N)^K$, independent
of the properties of the bosonic matter. Yet, we expect that the bosonic matter should be important: if two lattice boson models realize
different short-range entangled phases with the same symmetry (i.e. different SPT phases\cite{chen13}), then presumably the corresponding
gauge theories belong to distinct phases as well, by analogy with the 2D case\cite{levin12,cheng13}. Clearly, if we want to distinguish these different types
of 3D gauge theories, we must consider braiding processes with more than two loops.

{\it Three-loop braiding statistics.} For these reasons, we are naturally led to consider a braiding process involving two loops $\alpha, \beta$ which are
linked with a third ``base'' loop $\gamma$ (Fig.~\ref{fig_braiding}).
When the loop $\alpha$ sweeps around $\beta$ in a right-handed manner, it can acquire a statistical Berry phase which we will
denote by $\theta_{\alpha \beta, c}$ where $\phi_\gamma = \frac{2\pi}{N} c$ with $c$ being an integer vector. We use the notation $\theta_{\alpha \beta, c}$, rather than
$\theta_{\alpha \beta, \gamma}$ because $\theta$ is insensitive to
the charge attached to $\gamma$ and depends only on its flux $\phi_\gamma = \frac{2\pi}{N}  c$. Similarly, we will also consider an
exchange or half-braiding process in which two identical loops $\alpha$, which are linked with a base loop with flux $\frac{2\pi}{N}  c$, are braided
through one another and exchange places. The statistical phase associated with this exchange will be denoted by $\theta_{\alpha, c}$. Note that throughout this paper we assume the loops have Abelian statistics.

\begin{figure}
\includegraphics{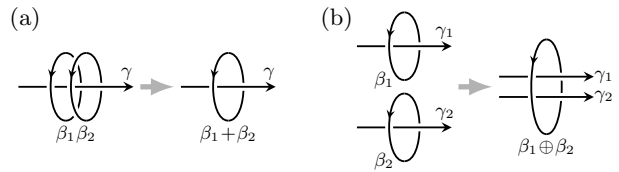}
\caption{Two ways to fuse loops together.}
\label{fig_fusion}
\end{figure}

These three-loop braiding processes are fundamentally different from the two-loop case because in the three-loop topology, the base loop $\gamma$
may prevent us from shrinking $\alpha$ and $\beta$ to a point and annihilating them locally. As a result,
the above argument that vortex loop statistics follow the Aharonov-Bohm law
(\ref{twostringstat}) is no longer valid. Thus, the three-loop braiding statistics are less constrained than the two-loop case.


{\it Constraints on $\theta_{\alpha \beta, c}$ and $\theta_{\alpha, c}$.}
We now discuss the basic physical constraints on the three-loop braiding statistics.
One of the simplest constraints is that
$\theta_{\alpha \beta, c} = \theta_{\beta \alpha, c}$. To derive this property,
we note that a process in which $\alpha$ winds around $\beta$ can be smoothly deformed into one in which $\beta$ winds around $\alpha$.
Therefore, since the statistical phase is invariant under smooth deformations of the braiding path, $\theta_{\alpha \beta, c}$ must
be symmetric in $\alpha$ and $\beta$. Another obvious constraint is that $\theta_{\alpha \alpha, c} = 2 \theta_{\alpha, c}$. This
relation is clear since a full braiding is equivalent to performing two exchanges in series.

\begin{figure}
\centering
\includegraphics{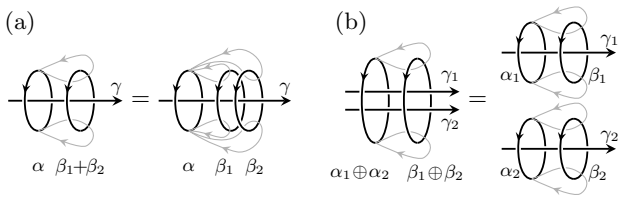}
\caption{Braiding processes associated with equations (\ref{linear2}) [panel (a)] and (\ref{linear3}) [panel (b)]. 
Here, $\phi_\gamma= \frac{2\pi}{N} c$, while $\phi_{\gamma_1}= \frac{2\pi}{N}  c_1$ and $\phi_{\gamma_2}= \frac{2\pi}{N} c_2$.}
\label{fig_linearity}
\end{figure}

Even more powerful constraints on $\theta$ can be obtained by thinking about ``fusion'' of
vortex loops. More specifically, there are two distinct ways to fuse loops together. In the first type of fusion process [Fig.~\ref{fig_fusion}(a)],
two loops $\beta_1, \beta_2$ that are linked to the same loop $\gamma$ can be fused to form a new loop `$\beta_1 + \beta_2$' that is also linked to
$\gamma$. In the second type of fusion process [Fig.~\ref{fig_fusion}(b)], two loops $\beta_1, \beta_2$ that share the same
flux $\phi_{\beta_1} = \phi_{\beta_2}$ but are linked with two different loops $\gamma_1$ and $\gamma_2$, can be fused to form a loop
`$\beta_1 \oplus \beta_2$', which is linked to both $\gamma_1$ and $\gamma_2$. It is not hard to see that $\theta_{\alpha \beta, c}$ must be
\emph{linear} under both fusion processes\cite{footnote2}  (Fig.~\ref{fig_linearity}):
\begin{eqnarray}
\theta_{\alpha (\beta_1+\beta_2), c} &=& \theta_{\alpha \beta_1, c} + \theta_{\alpha \beta_2,c}; \label{linear2} \\
\theta_{(\alpha_1 \oplus \alpha_2) (\beta_1 \oplus \beta_2),(c_1+c_2)} &=& \theta_{\alpha_1 \beta_1, c_1} +
\theta_{\alpha_2 \beta_2,c_2}; \label{linear3}
\end{eqnarray}
To derive these identities, it suffices to show that the processes defined by the left hand sides of Fig.~\ref{fig_linearity}a-b can be smoothly
deformed into the processes corresponding to the right hand sides of Fig.~\ref{fig_linearity}a-b. These deformations are described in the
Supplementary Material\cite{supplementary}.

One implication of the linearity of $\theta$ (\ref{linear2}-\ref{linear3}) is that we can reconstruct all the three-loop statistics
from the statistics of vortex loops with \emph{unit} flux. The statistics of these unit fluxes can in turn be
summarized by two tensors $\Theta_{ij,k}$ and $\Theta_{i,k}$. These tensors are defined by
\begin{align}
\Theta_{ij,k} \equiv N \theta_{\alpha \beta, e_k} , \ \ \ \ \ \ \ \Theta_{i,k} \equiv N \theta_{\alpha, e_k} \label{unitfluxtensor}
\end{align}
where $\alpha, \beta$ are \emph{any} two loops carrying unit flux $\phi_\alpha = \frac{2\pi}{N} e_i$ and
$\phi_\beta = \frac{2\pi}{N} e_j$ respectively, and where $e_i \equiv (0,\dots,1,\dots,0)$ with a $1$ in the $i$th entry and $0$
everywhere else.
To see why the tensor $\Theta_{ij,k}$ is well-defined modulo $2\pi$, note that if we choose another set of loops $\alpha', \beta'$ with
the same flux, then the only possible topological
difference between $\alpha', \beta'$ and $\alpha, \beta$ is that they may have different amounts of charge attached to them. But from
the Aharonov-Bohm formula (\ref{particlestat}), we see that attaching charge to $\alpha$ and $\beta$ can only shift the value of
$\theta_{\alpha \beta, e_k}$ by multiples of $2\pi/N$ and hence can only shift $\Theta_{ij,k}$ by multiples of $2\pi$. Similar reasoning
applies in the case of $\Theta_{i,k}$.

Given that the $\Theta_{ij,k}$ and $\Theta_{i,k}$ effectively summarize all the three-loop statistics,
all that remains is to find the physical constraints on these two quantities. We now argue that these
constraints are as follows:
\begin{align}
&\Theta_{ij,k}  =  \Theta_{ji,k}, \quad\quad\quad \ \ \ \ \Theta_{ii,k} = 2 \Theta_{i,k}, \label{mutex} \\
&\Theta_{ij,k} + \Theta_{jk,i}  +  \Theta_{ki,j}  = 0,  \label{cyclic}\\
&\Theta_{ik,i} + \Theta_{i,k}  = 0, \quad\ \ \ \ \ \ \Theta_{i,i}= 0,\label{cyclicex} \\
&\Theta_{ij,k}  = \frac{2\pi}{N}\cdot(\text{integer}),\ \ \Theta_{i,k} = \frac{2\pi}{N}\cdot (\text{integer}). \label{quant}
\end{align}
The first two constraints (\ref{mutex}) are obvious, since they are special cases of the more general relations discussed above.
The quantization conditions (\ref{quant}) are also easy to derive: for example, to prove the first equation in (\ref{quant}),
consider a thought experiment in which a loop $\alpha$ carrying flux $\phi_\alpha = \frac{2\pi}{N}  e_i$, together with $N$ identical loops
$\beta$ carrying flux $\phi_\beta = \frac{2\pi}{N}  e_j$,
are all linked to a common base loop with flux $\frac{2\pi}{N}  e_k$. Now imagine we fuse the $\beta$ loops together to form a
new loop $B$ and then we braid $\alpha$ around $B$. By the linearity of $\theta$ (\ref{linear2}),  the resulting statistical phase is
\begin{equation}
\theta_{\alpha B,e_k} = N  \theta_{\alpha \beta, e_k} = \Theta_{ij,k}
\end{equation}
At the same time, we can see that $\phi_B = N \phi_\beta = 0$ so $B$ is a pure charge. It then follows from the Aharonov-Bohm formula
(\ref{particlestat}) that $\theta_{\alpha B,e_k} = q_B \cdot \phi_\alpha$, which is a multiple of $2\pi/N$. Combining these two observations
we deduce that $\Theta_{ij,k}$ is a multiple of $2\pi/N$. The proof of the second equation in (\ref{quant}) is similar.

Equations (\ref{cyclic}-\ref{cyclicex}) are the most interesting constraints on $\Theta$ as these relations
have no analogues in the theory of 2D braiding statistics. We call Eq.~(\ref{cyclic}) the {\it cyclic} relation. A physical
derivation of the cyclic relation is given in the Supplementary Material\cite{supplementary}.
The first equation in (\ref{cyclicex}) can be proved in a similar manner. On the other hand, we do not have a physical derivation of
$\Theta_{i,i} = 0$, so this constraint on $\Theta$ is simply a conjecture. This conjecture is
supported by two pieces of evidence: first, all the microscopic models constructed below obey this relation. Second, we can
prove the weaker, but closely related relation $3 \Theta_{i,i} = 0$ (mod $2\pi$) using the second equation in (\ref{mutex}) together with the first
equation in (\ref{cyclicex}).

\begin{figure}
\centering
\includegraphics{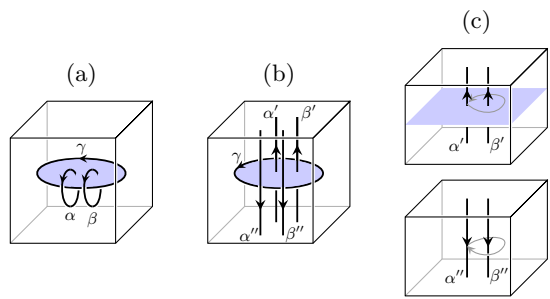}
\caption{Computing three-loop statistics from 2D braiding.}
\label{fig_dimreduction}
\end{figure}

{\it Dimensional reduction.} We now derive a formula for the three-loop statistics that will be useful in analyzing the microscopic models discussed later. This formula is obtained by considering our system in an $L_x \times L_y \times L_z$ torus geometry --- i.e. a geometry with periodic boundary conditions in all three directions.
Let $\alpha, \beta$ be two loops linked with a base loop $\gamma$ carrying flux $\phi_\gamma = \frac{2 \pi}{N}  c$ [Fig.~\ref{fig_dimreduction}(a)].
For concreteness, suppose that $\gamma$ lies in the $xy$ plane while $\alpha, \beta$ lie in the $xz$ plane. When $\alpha$ sweeps around $\beta$, it gives
rise to a statistical phase $\theta_{\alpha\beta,c}$ which we wish to compute. To this end, we stretch $\alpha$ in the $z$ direction until it wraps all
the way around the periodic $z$ direction. We can then fuse $\alpha$ with itself, thereby splitting $\alpha$ into two noncontractible loops $\alpha'$ and $\alpha''$
[Fig~\ref{fig_dimreduction}(b)]. Similarly, we can stretch $\beta$ in the $z$ direction and fuse it with itself so that it splits into $\beta'$ and $\beta''$.
It is clear that the braiding process involving $\alpha$ and $\beta$ can now be decomposed into two separate processes in which $\alpha'$ is braided
around $\beta'$ and $\alpha''$ is braided around $\beta''$. Since these two processes are separate, we can think of them as taking place in two separate systems [Fig.~\ref{fig_dimreduction}(c)]. Furthermore, for the process involving $\alpha', \beta'$ we can stretch $\gamma$ in the $xy$ plane so that it fuses and
annihilates with itself. This effectively leaves a gauge flux $\phi_\gamma = \frac{2 \pi}{N}  c$ through one of the three holes of the 3D
torus --- more precisely, the hole bounded by a noncontractible cycle along the $z$ direction (the ``$z$-hole''). Likewise, for the process involving
$\alpha'', \beta''$ we can shrink $\gamma$ in the $xy$ plane until it fuses and annihilates with itself, leaving no gauge flux through the ``$z$-hole.''
In this way, we see that $\theta_{\alpha \beta, c}$ can be expressed as
\begin{equation}
\theta_{\alpha \beta, c} = \theta_{\alpha' \beta',c} - \theta_{\alpha'' \beta'',0}
\label{3dtorus}
\end{equation}
where the quantities on the right hand side are statistical phases associated with braiding two vortex lines around one another in the $xy$ plane.
In the first term, this braiding takes place in the presence of a gauge flux $\frac{2\pi}{N}c$  through the ``$z$-hole'', while in the second
term there is no such gauge flux. The relative sign comes from the fact that the two pairs are braided in opposite directions. The formula (\ref{3dtorus}) is useful because each of the terms on the right hand side can be thought of as braiding statistics of a 2D system if we take the thermodynamic limit
$L_x, L_y \rightarrow \infty$, while keeping $L_z$  finite but larger than the correlation length. An analogous formula can be derived for the
exchange statistics $\theta_{\alpha,c}$.

{\it Microscopic models.} To obtain examples of systems with nontrivial three-loop statistics, we consider
``gauged'' SPT models --- that is, we take the exactly soluble lattice boson models of Ref. \onlinecite{chen13} which realize
different SPT phases, and we couple them to a gauge field. These gauged SPT models can be equivalently\cite{levin12} thought of as Dijkgraaf-Witten models\cite{dijkgraaf90}. As above, we focus on the case where the symmetry group is $G = (\mathbb Z_N)^K$.

As discussed in Ref.~\onlinecite{chen13}, the basic input for constructing a 3D gauged SPT model is a {\it $4$-cocycle} $\omega: G^4 \rightarrow U(1)$. If two cocycles $\omega_1, \omega_2$
differ by a {\it $4$-coboundary} $\nu$, i.e. $\omega_1 = \omega_2 \cdot \nu$, then the corresponding models belong to the same SPT phase.
Thus, inequivalent models are classified by elements of the cohomology group $H^4[G, U(1)]$.
Here we focus on $4$-cocycles $\omega$ of the form
\begin{align}
\omega(a,b,c,d) =  e^{\frac{i 2\pi}{N^2}\sum_{ijk}M_{ijk} a_ib_j(c_k+d_k-[c_k+d_k])},
\label{omega}
\end{align}
where $M_{ijk}$ is an integer tensor and we parameterize the different group elements of $G = (\mathbb Z_N)^K$ with
integer vectors $a= (a_1, \dots, a_K)$ with $a_i = 0,\dots,N-1$. The square bracket $[c_k+d_k]$ is defined to be $c_k+d_k$ (mod $N$)
with values taken in the range $0,\dots,N-1$.

Our task is to compute the three-loop statistics $\Theta_{ij,k}$ and $\Theta_{i,j}$ of the gauged SPT model with cocycle $\omega$. The details of this calculation, which is based on
the formula (\ref{3dtorus}), can be found in the Supplementary Material\cite{supplementary}. The end result is
\begin{align}
\Theta_{ij,k} & = \frac{2\pi}{N} (M_{ikj} - M_{kij} +  M_{jki} - M_{kji}), \nonumber \\
\Theta_{i,j} & = \frac{2\pi}{N }(M_{iji} - M_{jii}). \label{statformula}
\end{align}
As a consistency check, one can easily verify that these expressions satisfy conditions
(\ref{mutex}-\ref{quant}). Conversely, it is a straightforward mathematical exercise to check that every $\Theta_{ij,k}$ and
$\Theta_{i,j}$ that obeys (\ref{mutex}-\ref{quant}) can be written in the form (\ref{statformula}) for some $M_{ijk}$. Hence, every solution to
(\ref{mutex}-\ref{quant}) can be physically realized as a gauged SPT model.

{\it Examples.} The simplest example is $G = \mathbb Z_N$. In this case, $M$ has only one component $M_{111}$, and (\ref{statformula}) gives
trivial loop statistics, $\Theta_{ii,i} = \Theta_{i,i} = 0$, for any choice of $M$. This is a reasonable result since
$H^4[\mathbb{Z}_N, U(1)] = 0$, so all the SPT models with $G = \mathbb{Z}_N$ are equivalent to product states\cite{chen13}.

The simplest nontrivial example is given by $G = (\mathbb Z_N)^2$. In this case, if we choose $M_{211} = p_1$, $M_{122} = p_2$
and all other components vanishing, we obtain the three-loop statistics shown in Table~\ref{tab}. We can see that there are $N^2$ distinct types of statistics
that can be realized by the gauged SPT models with $G =(\mathbb Z_N)^2$. This is also a reasonable result since
$H^4[(\mathbb{Z}_N)^2, U(1)] = (\mathbb{Z}_N)^2$ so the SPT models realize $N^2$ distinct phases\cite{chen13, chen13b}. Evidently, each phase is associated with
a different type of three-loop statistics.

\begin{table}
\caption{$\Theta_{ij,k}$ for the SPT models with $(\mathbb Z_N)^2$ symmetry.}
\begin{tabular}{cccccc}
\hline  \hline
$\ \ \Theta_{11, 1}\ \ $ &  $\ \ \Theta_{12,1}\ \ $ & $\ \ \Theta_{22,1}\ \ $ & $\ \ \Theta_{11,2}\ \ $ & $\ \ \Theta_{12,2}\ \ $ &
$\ \ \Theta_{22,2}\ \ $\\
\hline
0 & $\frac{2\pi}{N} p_1$ & $-\frac{4\pi}{N} p_2$ & $-\frac{4\pi}{N} p_1$ &  $\frac{2\pi}{N} p_2$ & 0 \\
\hline
\label{tab}
\end{tabular}
\end{table}

{\it Discussion.} The above examples show that the gauged SPT phases with $G = \mathbb{Z}_N$ and $G =(\mathbb Z_N)^2$ are uniquely
characterized by their three-loop statistics. More generally, we find it plausible that
every 3D SPT phase with unitary symmetries is uniquely characterized by its three-loop statistics --- similarly to what has been
proposed in the 2D case\cite{levin12}. One subtlety in checking this conjecture for more general $G = (\mathbb{Z}_N)^K$ is that when $K \geq 4$, the cocycles
(\ref{omega}) do not exhaust all elements of $H^4[G,U(1)]$. Furthermore, the remaining elements of $H^4[G,U(1)]$ can lead to non-Abelian three-loop statistics (see Ref. \onlinecite{propitius95} for examples of this phenomenon in the 2D case). Thus, a theory of non-Abelian loop statistics may be necessary to proceed further in this direction.

Is three-loop statistics measurable? In principle, three-loop statistics could be measured experimentally by performing interferometry on loop-like excitations; in practice, such an experiment would be challenging. A more straightforward application is to numerical simulations, where three-loop statistics could be directly extracted from an appropriate Berry phase computation.

After submitting this paper for publication, we became aware of an independent work\cite{ran14} containing related results on three-loop statistics. Other recent work on this topic includes Refs.\cite{jwang14, jian14}.  We thank M. Cheng, C.-H. Lin and A. Vishwanath for helpful discussions. This work is supported by the Alfred P. Sloan foundation and NSF under grant No. DMR-1254721.

\clearpage


\appendix

\section{Supplementary Material}
This supplementary material contains four parts. In the first two parts, we prove that the three-loop statistics obeys two relations, Eq.~(\ref{linear3}) and
Eq.~(\ref{cyclic}), from the main text. In the third part, we review the basics of group cohomology and Dijkgraaf-Witten models. In the last part,
we derive Eq.~(\ref{statformula}) from the main text, which gives an explicit formula for the three-loop statistics in the Dijkgraaf-Witten models with group $G = (\mathbb Z_N)^K$.

\subsection{1. Proof of the linearity relation (4)}
\begin{figure}[b]
\centering
\includegraphics{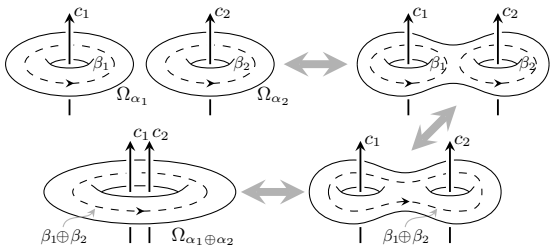}
\caption{A smooth deformation between the braiding processes associated with the two sides of Eq.~(\ref{linear3}). We use
$\Omega_{\alpha_1}$, $\Omega_{\alpha_2}$, and $\Omega_{\alpha_1\oplus\alpha_2}$ to denote the surfaces swept out by
$\alpha_1$, $\alpha_2$ and $\alpha_1\oplus\alpha_2$, respectively. The dashed lines denote the vortex loops $\beta_1$, $\beta_2$ and
$\beta_1 \oplus \beta_2$ which are enclosed by these surfaces. }
\label{fig_topoequiv}
\end{figure}
In this section, we prove Eq.~(\ref{linear3}), which we reprint below for convenience:
\begin{displaymath}
\text{Eq.~(\ref{linear3})}: \ \theta_{(\alpha_1 \oplus \alpha_2) (\beta_1 \oplus \beta_2),(c_1+c_2)} = \theta_{\alpha_1 \beta_1, c_1} + \theta_{\alpha_2 \beta_2,c_2}.
\end{displaymath}
This relation states that $\theta$ is linear under one of the two types of fusion processes for loops. We will not present the proof of
the other linearity relation, Eq.~(\ref{linear2}), since Eq.~(\ref{linear2}) can be established in the same way as the familiar result
that 2D Abelian mutual statistics $\theta_{\alpha \beta}$ is linear under fusion of quasiparticles:
$\theta_{\alpha (\beta_1 + \beta_2)} = \theta_{\alpha \beta_1} + \theta_{\alpha \beta_2}$. Our focus is on the ``more 3D'' relation
(\ref{linear3}).

To prove Eq.~(\ref{linear3}), it is enough to show that the braiding process associated with one side of Eq.~(\ref{linear3}) can be smoothly
deformed to the process associated with the other side of Eq.~(\ref{linear3}). Then, since the statistical phase is invariant under ``smooth''
deformations\cite{footnote3} of the braiding path, the relation will follow
immediately. The desired deformation is shown in Fig.~\ref{fig_topoequiv}. The lower-left panel of Fig.~\ref{fig_topoequiv} shows the torus
$\Omega_{\alpha_1 \oplus \alpha_2}$ which is swept out by the loop $\alpha_1\oplus\alpha_2$ in the braiding process associated with
$\theta_{(\alpha_1 \oplus \alpha_2) (\beta_1 \oplus \beta_2),(c_1+c_2)}$. Similarly, the upper-left panel of Fig.~\ref{fig_topoequiv} shows
the two tori $\Omega_{\alpha_1}$ and $\Omega_{\alpha_2}$ swept out by the loops $\alpha_1$ and $\alpha_2$ in the braiding processes associated
with $\theta_{\alpha_1 \beta_1, c_1}$ and $\theta_{\alpha_2 \beta_2,c_2}$. The two right panels show the middle steps of the deformation.

\subsection{2. Derivation of the cyclic relation (7)}

\begin{figure}[t]
\centering
\includegraphics{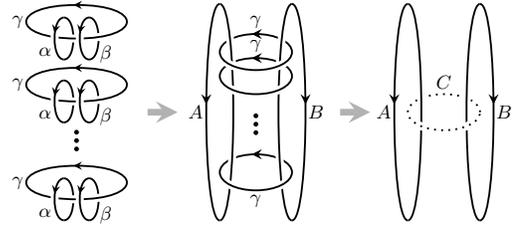}
\caption{First thought experiment to prove the cyclic relation (\ref{cyclic}). The loop $A$ is obtained by fusing $N$ identical loops
$\alpha$, that is, $A=\alpha\oplus\cdots\oplus\alpha$. Similarly, $B=\beta\oplus\cdots\oplus\beta$ and $C=\gamma+\dots+\gamma$. }
\label{fig_firstthought}
\end{figure}

\begin{figure}[b]
\centering
\includegraphics{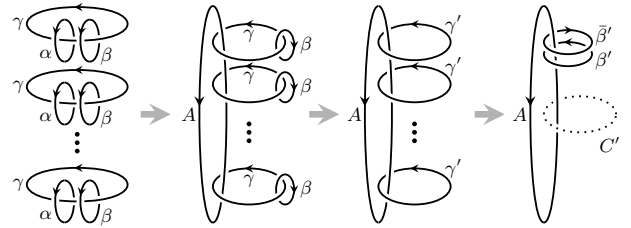}
\caption{Second thought experiment to prove the cyclic relation (\ref{cyclic}).}
\label{fig_secondthought}
\end{figure}

\begin{figure}[t]
\centering
\includegraphics{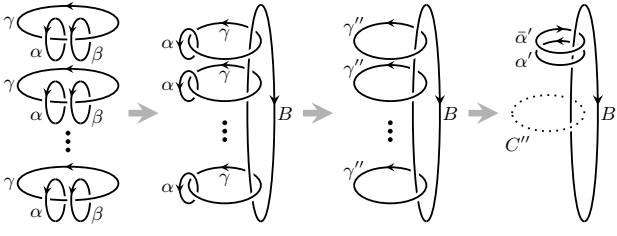}
\caption{Third thought experiment to prove the cyclic relation (\ref{cyclic}).}
\label{fig_thirdthought}
\end{figure}

In this section, we prove Eq.~(\ref{cyclic}), which we reprint below for convenience:
\begin{displaymath}
\text{Eq.~(\ref{cyclic})}: \  \Theta_{ij,k} + \Theta_{jk,i} + \Theta_{ki,j} = 0.
\end{displaymath}
To prove Eq.~(\ref{cyclic}), we consider a series of thought experiments, shown in Figs.~\ref{fig_firstthought}, \ref{fig_secondthought} and \ref{fig_thirdthought}. In the first thought experiment, we create $N$ identical links $\{\alpha, \beta, \gamma\}$,
carrying flux $\phi_\alpha = \frac{2\pi}{N} e_i$, $\phi_\beta = \frac{2\pi}{N} e_j$ and $\phi_\gamma = \frac{2\pi}{N} e_k$ respectively (Fig.~\ref{fig_firstthought}). We then imagine fusing the $N$ links together to form a single
link made up of three loops, $A,B,C$ where $\phi_A = \frac{2\pi}{N} e_i$, $\phi_B = \frac{2\pi}{N} e_j$ and $\phi_C = N\cdot \frac{2\pi}{N} e_k = 0 \ ({\rm mod} \ 2\pi)$. After the fusion, we imagine braiding $A$ around $B$ with the base $C$. Given the linearity of $\theta$ (\ref{linear3}), the statistical phase associated with this braiding is
\begin{equation}
\theta_{AB,N e_k} = N \theta_{\alpha \beta, e_k} = \Theta_{ij,k}.
\end{equation}
At the same time, since $\phi_C=0$, the braiding between $A,B$ is no different from two-loop braiding. Therefore, we can use (\ref{twostringstat}) to deduce
$\theta_{AB,N e_k} = q_{A} \cdot \phi_B + q_{B} \cdot \phi_A$. We conclude that
\begin{eqnarray}
\Theta_{ij,k} &=& q_{A} \cdot \phi_B + q_{B} \cdot \phi_A \nonumber \\
&=& \frac{2\pi}{N} \left(q_{A} \cdot e_j + q_{B} \cdot e_i\right).
\label{abc}
\end{eqnarray}

We now derive similar expressions for $\Theta_{jk,i}$ and
$\Theta_{ki,j}$. In the case of $\Theta_{jk,i}$, we imagine another thought experiment (Fig.~\ref{fig_secondthought}), where we fuse the $\alpha$ loops
together to form $A$, but we don't fuse the $\beta$ or $\gamma$ loops. We then shrink the $\beta$ loops and fuse the
$\beta$ onto $\gamma$ to form composite loops $\gamma' = \gamma \cup \beta$. Next we fuse the $\gamma'$ loops to
form a new loop $C'=\gamma'+\cdots+\gamma'$. Finally, at the end, we create a vortex-antivortex pair of loops $\beta'$ and $\bar\beta'$, with $\beta'$ carrying unit flux
$\phi_{\beta'} = \frac{2\pi}{N} e_j$, and we imagine braiding $\beta'$ around $C'$.  Given
the linearity of $\theta$ (\ref{linear2}), the statistical phase associated with this braiding process is
\begin{equation}
\theta_{\beta' C', e_i} = N \theta_{\beta' \gamma', e_i} = \Theta_{jk,i}.
\end{equation}
At the same time, since $\phi_{C'} = N \cdot \frac{2\pi}{N} e_k = 0  \ ({\rm mod} \ 2\pi)$, we see that $C'$ is a pure charge so we can use (\ref{particlestat}) to write $\theta_{\beta' C',e_i} = q_{C'} \cdot \phi_{\beta'}$. Hence
\begin{eqnarray}
\Theta_{jk,i} &=& q_{C'} \cdot \phi_{\beta'} \nonumber \\
&=& \frac{2\pi}{N} q_{C'} \cdot e_j.
\label{bca}
\end{eqnarray}

Finally, to compute $\theta_{ki,j}$, we consider a third thought experiment (Fig.~\ref{fig_thirdthought}), where we fuse the $\beta$ loops together to form $B$, but we don't fuse the $\alpha$ or $\gamma$. The composite loops $\gamma'' = \gamma \cup \alpha$ are then fused together to form $C''=\gamma''+\cdots+\gamma''$. At the end, we create a vortex-antivortex pair of loops $\alpha'$ and $\bar\alpha'$, with $\alpha'$ carrying unit flux $\phi_{\alpha'} = \frac{2\pi}{N} e_i$, and we imagine braiding $C''$ around $\alpha'$. By the same reasoning as above, we have
\begin{equation}
\theta_{C'' \alpha', e_j} = \Theta_{ki,j} = \frac{2\pi}{N} q_{C''} \cdot e_i.
\label{cab}
\end{equation}
To complete the derivation, we note that
\begin{equation}
\frac{2\pi}{N}(q_{A} + q_{C'}) \cdot e_j = 0, \ \ \  \frac{2\pi}{N}(q_{B} + q_{C''}) \cdot e_i = 0.
\label{neutrality}
\end{equation}
Here the first relation follows from the observation that $(q_A + q_{C'}) = q_{\text{tot}}$ is the total charge on the $N$ identical links and is
therefore divisible by $N$; the second relation follows by the same reasoning. Adding together (\ref{abc}), (\ref{bca}), (\ref{cab}), and using
(\ref{neutrality}), we derive the cyclic relation (\ref{cyclic}).

\subsection{3. Group cohomology and Dijkgraaf-Witten models}
In this section, we give a brief review of the basics of group cohomology and Dijkgraaf-Witten models. The purpose of this review is to provide the reader with the necessary background to understand the next section where we will compute the three-loop statistics in (3+1)D Dijkgraaf-Witten models.

\subsubsection{Basics of group cohomology}
In the following, we review the basic elements of the cohomology of finite groups\cite{chen13, propitius95, brown}. We focus on the
cohomology group $H^n[G,U(1)]$, as only this case is relevant for this paper.

Let $G$ be a finite group. The basic objects that group cohomology studies are \emph{$n$-cochains}. An $n$-cochain is a $U(1)$ valued function $c(g_1, \dots, g_n)$:
\begin{displaymath}
c: \underbrace{G\times G\times \cdots \times G}_{n \text{ times}} \rightarrow U(1).
\end{displaymath}
The collection of $n$-cochains form an Abelian group $\mathcal C^n$, where the group operation is defined by
\begin{displaymath}
(c_1\cdot c_2)(g_1, \dots, g_n) = c_1(g_1, \dots, g_n)\cdot c_2(g_1, \dots, g_n).
\end{displaymath}
The \emph{coboundary operator} $\delta$ is a map $\delta: \mathcal C^n\rightarrow \mathcal C^{n+1}$, defined by
\begin{align}
\delta c &(g_1, \dots, g_{n+1}) = c(g_2, \dots, g_{n+1})c(g_1, \dots, g_n)^{(-1)^{n+1}} \nonumber\\
& \times \prod_{i=1}^n [c(g_1, \dots, g_i g_{i+1}, \dots, g_{n+1})]^{(-1)^{i}}. \label{delta}
\end{align}
It is easy to check that the coboundary operator satisfies $\delta(c_1\cdot c_2) = \delta c_1 \cdot \delta c_2$. More importantly, one can check that
$\delta$ is nilpotent: $\delta^2=1$.

With the help of the coboundary operator, we can now define \emph{$n$-cocycles} and \emph{$n$-coboundaries}. An $n$-cocycle is an $n$-cochain $\omega$ that
satisfies $\delta \omega =1$. Likewise, an $n$-coboundary is an $n$-cochain $\nu$ that can be written as $\nu =\delta  c$ where $c\in\mathcal C^{n-1}$.
The nilpotence of $\delta$ implies that a coboundary must also be a cocycle. This allows us to define an equivalence relation for the cocycles: two $n$-cocycles
$\omega_1$ and $\omega_2$ are said to be {\it cohomologically equivalent} if and only if $\omega_1 = \omega_2 \cdot \delta c$, for some $c\in \mathcal C^{n-1}$.
The equivalence classes of the $n$-cocycles form an Abelian group, called the $n$th \emph{cohomology group}, which is denoted by $H^{n}[G, U(1)]$.

\subsubsection{Dijkgraaf-Witten models}
The Dijkgraaf-Witten models\cite{dijkgraaf90} are exactly soluble lattice models that realize different types of lattice gauge theories with finite gauge group. Here we review the space-time path integral formulation of these models. For a Hamiltonian formulation of these models in (2+1)D, see e.g. Ref.~\cite{hu13}.

The basic input needed to construct a $d$-dimensional Dijkgraaf-Witten model with gauge group $G$ is (1) a $d$-cocycle $\omega$ and (2)
a triangulation of $d$-dimensional Euclidean space-time. To build the associated Dijkgraaf-Witten model, we label the vertices of the triangulation in an
ordered sequence $\{p,q,r, \dots\}$. The degrees of freedom in the model are group elements $h_{pq} \in G$ which live on the edges $[pq]$
of the triangulation, and can be thought of as gauge fields. For each gauge field configuration $\{h_{pq}\}$, the corresponding action $e^{-S(\{h_{pq}\})}$ is defined by the following recipe.
First, one needs to determine if the configuration $\{h_{pq}\}$ is \emph{flat}, that is $h_{pq} h_{qr} h_{rp} = 1$ for every $2$-simplex $[pqr]$.
If $\{h_{pq}\}$ is not flat, then $e^{-S(\{h_{pq}\})} = 0$. On the other hand, if $\{h_{pq}\}$ is flat, then $e^{-S(\{h_{pq}\})}$ is given by a
product of complex weights, one for every $d$-dimensional simplex in the triangulation. For example, the action for a (3+1)D Dijkgraaf-Witten model with $4$-cocycle
$\omega$ is given by
\begin{equation}
e^{-S(\{h_{pq}\})} =  \prod_{[pqrst]} \left[\omega(h_{pq}, h_{qr}, h_{rs}, h_{st})\right]^{\sigma_{pqrst}}.
\end{equation}
Here $[pqrst]$ ($p<q<r<s<t$) runs over the $4$-simplices in the triangulation and $\sigma_{pqrst}=\pm 1$ is a chirality factor which is determined by the ordering of the vertices in
each $4$-simplex. The action for Dijkgraaf-Witten models in other dimensions is similar.

To obtain the partition function, we sum over gauge field configurations and multiply by a normalization factor of $\frac{1}{|G|^{N_v}}$ where $|G|$ is the number of group elements in $G$ and $N_v$ is the number of vertices in the triangulation. For example, in the (3+1)D case, the partition function is
\begin{eqnarray}
Z &=& \frac{1}{|G|^{N_v}}\sideset{}{'}\sum_{\{h_{pq}\}} e^{-S(\{h_{pq}\})} \nonumber \\
&=& \frac{1}{|G|^{N_v}}\sideset{}{'}\sum_{\{h_{pq}\}} \prod_{[pqrst]} \left[\omega(h_{pq}, h_{qr}, h_{rs}, h_{st})\right]^{\sigma_{pqrst}},\ \ \label{supp-dw}
\end{eqnarray}
where the summation $\sum'$ is taken over flat gauge field configurations only, since $e^{-S(\{h_{pq}\})} = 0$ for the other configurations.

It can be shown that if $\omega_1, \omega_2$ belong to the same cohomological equivalence class, then the corresponding Dijkgraaf-Witten models are equivalent, i.e. share the same partition function. Thus, the inequivalent $d$-dimensional Dijkgraaf-Witten models are classified by the cohomology group $H^{d}[G, U(1)]$.

\subsection{4. Loop statistics of (3+1)D Dijkgraaf-Witten models/gauged SPT models}
In this section, we derive the formula (\ref{statformula}) from the main text. This formula
gives the three-loop statistics of (3+1)D gauged SPT models with symmetry group
$G = (\mathbb Z_N)^K$. Alternatively the formula (\ref{statformula}) can be thought of as giving the three-loop statistics
of Dijkgraaf-Witten models with $G = (\mathbb Z_N)^K$, since gauged SPT models
are exactly equivalent to Dijkgraaf-Witten models. (See Ref.~\onlinecite{levin12} for a
discussion of this equivalence in (2+1) dimensions). For notational reasons, we find it
more convenient  use the Dijkgraaf-Witten language, so in what follows we will
phrase our calculation in terms of the three-loop braiding statistics of Dijkgraaf-Witten models.

Our derivation of (\ref{statformula}) proceeds in three steps. In the first step,
we derive a ``dimensional reduction'' formula that relates the vortex loop statistics of
(3+1)D Dijkgraaf-Witten models to the vortex statistics of (2+1)D Dijkgraaf-Witten
models. In the second step, we review some previously known results on vortex statistics in (2+1)D Dijkgraaf-Witten models. In the final step, we put everything together and we derive the formula (\ref{statformula}).

\subsubsection{Dimensional reduction of (3+1)D Dijkgraaf-Witten models}

In this section, we derive a ``dimensional reduction'' formula (\ref{supp-dimred})
that relates the vortex loop statistics of (3+1)D Dijkgraaf-Witten models to the vortex statistics of (2+1)D Dijkgraaf-Witten models.
As in the main text, we assume  $G = (\mathbb{Z}_N)^K$ and we restrict our analysis to models with Abelian three-loop statistics.

The starting point for our derivation is Eq. (\ref{3dtorus}) from the main text, which we repeat below for convenience:
\begin{equation}
\theta_{\alpha \beta, c} = \theta_{\alpha' \beta',c} - \theta_{\alpha'' \beta'',0}. \label{supp-3dtorus}
\end{equation}
This relation allows us to compute the three-loop statistics $\theta_{\alpha \beta, c}$ using a 3D spatial torus geometry.
Here, the quantity
$\theta_{\alpha'\beta',c}$ is the mutual statistics between two noncontractible vortex lines $\alpha'$ and
$\beta'$ oriented along the $z$ direction of a 3D torus, in the presence of a gauge flux $\frac{2\pi}{N}c$
through the ``$z$-hole'' of the 3D torus
(Fig.~\ref{fig_dimreduction}). Similarly, the quantity $\theta_{\alpha''\beta'',0}$ is the mutual statistics between
noncontractible vortex lines $\alpha''$ and $\beta''$ in the absence of a gauge flux through the ``$z$-hole'' of
the 3D torus.

\begin{figure}
\centering
\includegraphics{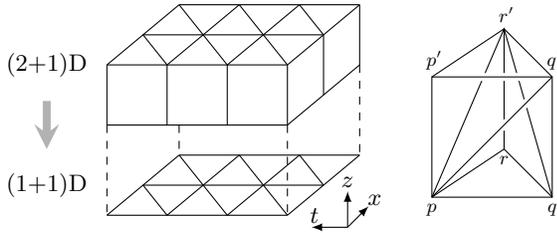}
\caption{Illustration of a (2+1)D-to-(1+1)D dimensional reduction for Dijkgraaf-Witten models. The (2+1)D space-time is triangulated
using a collection of triangular prisms and the periodic $z$ direction contains one lattice constant only.
Each triangular prism $[pqrp'q'r']$ contains three tetrahedra $[pqrr']$, $[pqq'r']$ and $[pp'q'r']$. The (2+1)D weight associated with the prism $[pqrp'q'r']$ can be rewritten as a (1+1)D weight of the corresponding triangle $[pqr]$. }
\label{fig_dimreduc_dw}
\end{figure}

Equation (\ref{supp-3dtorus}) is especially useful for analyzing Dijkgraaf-Witten models since these models
have a vanishing correlation length, and hence we can use a 3D torus of any size and triangulation that we like. Here we find it convenient
to choose a ``thin torus'', i.e. a 3D spatial torus with a thickness of only one unit cell in the $z$ direction and
arbitrary dimensions in the $x$ and $y$ directions (see Fig.~\ref{fig_dimreduc_dw} for an analogous geometry in one lower dimension).
Then, since our (3+1)D system has finite thickness in the $z$ direction, we can think of it as a (2+1)D lattice model. In fact,
we will show below that if we fix the gauge flux through the $z$-hole of the torus to be some group element $h \in G$, then this (2+1)D lattice model is
exactly the (2+1)D Dijkgraaf-Witten model associated with the 3-cocycle
\begin{equation}
\chi_h(h_1, h_2, h_3) = \frac{\omega(h_1, h_2, h_3, h)\omega(h_1, h, h_2, h_3)}{\omega(h_1, h_2, h, h_3)\omega(h, h_1, h_2, h_3)},\label{supp-c}
\end{equation}
where $\omega$ is the 4-cocycle associated with the (3+1)D Dijkgraaf-Witten model. Given the above result, we can rewrite Eq. (\ref{supp-3dtorus}) as
\begin{equation}
\theta_{\alpha \beta, c} = \theta^{2D}_{\alpha' \beta'}(c) - \theta^{2D}_{\alpha'' \beta''}(0) \label{supp-dimred}
\end{equation}
where $\theta^{2D}_{\alpha' \beta'}(c)$ denotes the mutual statistics of two vortex excitations $\alpha', \beta'$ in the
(2+1)D Dijkgraaf-Witten model (\ref{supp-c}) with 3-cocycle $\chi_c$, while $\theta^{2D}_{\alpha'' \beta''}(0)$ denotes the braiding
statistics of two vortex excitations $\alpha'', \beta''$ in the (2+1)D Dijkgraaf-Witten model (\ref{supp-c}) with 3-cocycle $\chi_0$.
Equation (\ref{supp-dimred}) is the desired dimensional reduction formula, and the main result of this section.

To complete our derivation, we now prove our earlier claim that the (3+1)D Dijkgraaf-Witten
model in the ``thin torus'' geometry is equivalent to a (2+1)D Dijkgraaf-Witten model with cocycle $\chi_h$.
In the first step, we consider a (3+1)D space-time geometry which is periodic in all spatial dimensions. We triangulate the (3+1)D space-time using a collection of 4D triangular prisms
with a thickness of one lattice constant along the $z$ spatial dimension; each 4D prism is further triangulated into four 4-simplices.
(An analogous triangulation in one lower dimension is shown in Fig.~\ref{fig_dimreduc_dw}). In this
triangulation, the action $e^{-S(\{h_{pq}\})}$ in (\ref{supp-dw}) can be written as a product of the local complex weights $W_{pqrsp'q'r's'}$ associated with each 4D prism $[pqrsp'q'r's']$. Here, we use the notation that all the links parallel to the $z$-axis are labeled by $[pp'], [qq'],\dots$, similarly to that in Fig.~\ref{fig_dimreduc_dw}. Each 4D prism $[pqrsp'q'r's']$ then contains four 4-simplices,
$[pqrss']$, $[pqrr's']$, $[pqq'r's']$ and $[pp'q'r's']$, so the local weight $W_{pqrsp'q'r's'}$ is given by
\begin{align}
&W_{pqrsp'q'r's'} = \nonumber\\
&\left[\frac{\omega(h_{pq}, h_{pr}, h_{rs}, h_{ss'})\omega(h_{pq}, h_{qq'}, h_{q'r'}, h_{r's'})}{\omega(h_{pq}, h_{qr}, h_{rr'}, h_{r's'})\omega(h_{pp'}, h_{p'q'}, h_{q'r'}, h_{r's'})}\right]^{\sigma_{pqrs}}
\label{suppW}
\end{align}
where $\sigma_{pqrs}$ is the chirality of the 3-simplex $[pqrs]$.

We next fix the gauge fields living on all $z$-links $[pp']$ to be $h_{pp'}=h\in G$. Different choices of $h$ will
correspond to different gauge fluxes through the ``$z$-hole'' of the 3D torus. Substituting $h_{pp'} = h$ into (\ref{suppW})
and remembering that periodicity along the $z$ dimension identifies the field $h_{pq}$ with $h_{p'q'}$, we obtain
$W_{pqrsp'q'r's'} = [\chi_h(h_{pq},h_{qr},h_{rs})]^{\sigma_{pqrs}}$ where
\begin{align}
\chi_h(h_{pq},  h_{qr}, h_{rs}) = \frac{\omega(h_{pq}, h_{qr}, h_{rs}, h)\omega(h_{pq}, h, h_{qr}, h_{rs})}{\omega(h_{pq}, h_{qr}, h, h_{rs})\omega(h, h_{pq}, h_{qr}, h_{rs})}. \label{supp-ch}
\end{align}
One may check that $\chi_h$ is a 3-cocycle for each $h$. The (3+1)D Dijkgraaf-Witten partition function (\ref{supp-dw}) then reduces to
\begin{equation}
Z_h =\frac{1}{|G|^{N_v}} \sideset{}{'} \sum_{\{h_{pq}\}} \ \prod_{[pqrs]} \left[\chi_h(h_{pq}, h_{qr}, h_{rs})\right]^{\sigma_{pqrs}}\label{supp-dw-2d},
\end{equation}
which is exactly the partition function of a (2+1)D Dijkgraaf-Witten model with cocycle $\chi_h$.

\subsubsection{Braiding statistics of (2+1)D Dijkgraaf-Witten models}
Here, we summarize some previously known results regarding the braiding statistics of (2+1)D Dijkgraaf-Witten models.
We list these results without proof; readers who are interested in how to obtain these results may consult
Refs.~\onlinecite{dijkgraaf90, propitius95, lin14}.

For simplicity, we focus on Dijkgraaf-Witten models with gauge group $G = (\mathbb Z_{N})^K$ and
Abelian quasiparticle statistics. It is known that the most general model with these properties can be constructed
from a 3-cocycle of the form\cite{propitius95}
\begin{equation}
\omega(a,b,c) =e^{\frac{i2\pi}{N^2} \sum_{ij} P_{ij}a_i(b_j+c_j-[b_j+c_j]) }, \label{supp-3cocyl}
\end{equation}
where $P$ is an arbitrary integer $K \times K$ matrix. Here we parameterize the different elements
of $G$ using integer $K$-component vectors $a=(a_1, \dots, a_K)$ with $a_i=0,\dots, N-1$, and the square bracket $[b_j+c_j]$ is defined to be
$b_j+c_j\ ({\rm mod} \ N)$ with values taken in the range $0, \dots, N-1$. It is easy to check that the above expression for
$\omega$ is indeed a 3-cocycle for any choice of $P$.

The most general quasiparticle excitations in the above models carry both flux and charge. In order to simplify our notation, we will call all of these excitations ``vortices''; pure charges will be regarded as special kinds of vortices carrying zero flux. General vortices can be labeled by ordered pairs $\alpha = (a,m)$ where $a,m$ are $K$-component integer vectors, $a = (a_1, \dots, a_K)$ and $m = (m_1, \dots, m_K)$ with $0 \le a_i, m_i \leq N-1$.
The label $a$ can be thought of as the amount of flux $\phi_\alpha =\frac{2\pi}{N} a$ carried by the vortex $\alpha$
while the label $m$ describes the amount of charge attached to $\alpha$. One subtlety is that the above labeling scheme is only
well-defined once we choose some conventions --- in particular, we have to pick a convention for which vortices are labeled as ``pure''
fluxes $(a,0)$. For concreteness, we will use the labeling convention defined in Ref.~\onlinecite{lin14} throughout our discussion.



The braiding statistics of the vortices are known and can be written down explicitly in terms of the matrix $P$. Specifically, the mutual
statistics between two vortices $\alpha = (a,m)$ and $\beta = (b,n)$ is given by
\begin{align}
\theta^{2D}_{\alpha\beta}=   \frac{2\pi}{N^2} \sum_{ij} (P_{ij}+P_{ji})a_ib_j +  \frac{2\pi}{N}  \sum_i (m_ib_i+n_ia_i),
\label{supp-2dmut}
\end{align}
while the exchange statistics of $\alpha = (a,m)$ is given by
\begin{align}
\theta^{2D}_\alpha = \frac{2\pi}{N^2} \sum_{ij} P_{ij}a_ia_j +  \frac{2\pi}{N}  \sum_i m_ia_i.
\end{align}
For a derivation of the above formulas, see e.g. Refs.~\onlinecite{propitius95,lin14}.



\subsubsection{Explicit formula for three-loop statistics}
We now derive the formula, Eq.~(\ref{statformula}), from the main text. This formula gives the three-loop statistics
for any $(3+1)D$ Dijkgraaf-Witten model with gauge group $G = (\mathbb Z_N)^K$ and with a $4$-cocycle of the form
\begin{align}
\omega(a,b,c,d) = &e^{ \frac{i2\pi}{N^2} \sum_{ijk}M_{ijk} a_ib_j(c_k+d_k-[c_k+d_k])}, \label{supp-4cocycle}
\end{align}
where $M$ is a three-index integer tensor. (The reason that we focus on the above class of $4$-cocycles is that
we believe that they are the most general cocycles such that the corresponding Dijkgraaf-Witten models have Abelian
loop statistics).

In the first step, we compute the ``dimensionally reduced'' $3$-cocycle
$\chi_h(a,b,c)$ (\ref{supp-c}) corresponding to $\omega$. Inserting the expression (\ref{supp-4cocycle}) into
(\ref{supp-c}), we find
\begin{equation}
\chi_h(a,b,c) =e^{\frac{i2\pi}{N^2} \sum_{ij} P_{ij}^h a_i(b_j+c_j-[b_j+c_j])},
\end{equation}
where $P^h$ is an integer matrix whose elements are given by
\begin{equation}
P_{ij}^h = \sum_k (M_{ikj} - M_{kij})h_k.
\end{equation}

Next, we note that the above 3-cocycle $\chi_h$ falls into the form (\ref{supp-3cocyl}), so we can immediately
write down the braiding statistics of the vortices in the (2+1)D Dijkgraaf-Witten models with 3-cocycle $\chi_h$.
In particular, according to (\ref{supp-2dmut}), the mutual statistics between two vortices $\alpha = (a,m)$ and
$\beta = (b,n)$ is given by
\begin{align}
\theta^{2D}_{\alpha\beta}(h) & =\frac{ 2\pi}{N^2}  \sum_{ij}(P_{ij}^h+P_{ji}^h) a_ib_j+\frac{2\pi}{N} \sum_i (m_i b_i+n_i a_i)
\label{supp-2dmut2}
\end{align}

In the final step, we use the dimensional reduction formula (\ref{supp-dimred}) to relate the three-loop statistics in
the (3+1)D Dijkgraaf-Witten model with cocycle (\ref{supp-4cocycle}) to the vortex statistics in the (2+1)D Dijkgraaf-Witten
models with cocycles $\chi_c$ and $\chi_0$:
\begin{equation}
\theta_{\alpha \beta, c} = \theta^{2D}_{\alpha' \beta'}(c) - \theta^{2D}_{\alpha''\beta''}(0) \label{supp-eq2d}
\end{equation}
Examining Fig.~\ref{fig_dimreduction}, it is clear that $\alpha, \alpha'$ carry the same flux as one another, while $\alpha''$ carries opposite flux. The same is true
for $\beta, \beta', \beta''$. Therefore, we can label these vortices as
\begin{align}
\alpha = (a,m), \ \alpha' = (a,m'), \ \alpha'' = (-a, m'') \nonumber \\
\beta = (b,n), \ \beta' = (b,n'), \ \beta'' = (-b, n'')
\end{align}
Substituting these expressions into (\ref{supp-2dmut2}) and using (\ref{supp-eq2d}), we derive
\begin{eqnarray}
\theta_{\alpha \beta, c} &=& \frac{ 2\pi}{N^2}  \sum_{ij}(P_{ij}^c+P_{ji}^c - P_{ij}^0-P_{ji}^0 ) a_ib_j \nonumber \\
		         &+& \frac{2\pi}{N} \sum_i ((m'_i+m''_i) b_i+(n'_i+n''_i) a_i)
\label{supp-3loop}
\end{eqnarray}
To get the unit flux statistics $\Theta_{ij,k}$ (\ref{unitfluxtensor}), we specialize to the case where $\alpha, \beta$ and $c$ carry unit flux,
that is $a = e_i$, $b= e_j$, $c = e_k$, and we multiply Eq. (\ref{supp-3loop}) by $N$. After these manipulations, the second term on the right
hand side drops out and we obtain
\begin{eqnarray}
N \theta_{\alpha \beta, e_k} &=& \frac{2\pi}{N }(P_{ij}^{e_k} + P_{ji}^{e_k} - P_{ij}^0-P_{ji}^0) \nonumber \\
&=& \frac{2\pi}{N }(M_{ikj} - M_{kij} + M_{jki} - M_{kji}).
\end{eqnarray}
We conclude that
\begin{equation}
\Theta_{ij,k} = N \theta_{\alpha \beta, e_k} = \frac{2\pi}{N }(M_{ikj} - M_{kij} + M_{jki} - M_{kji}).
\end{equation}
The exchange statistics $\Theta_{i,j}$ can be computed in a similar way, and is given by
\begin{equation}
\Theta_{i,j} = \frac{2\pi}{N }(M_{iji} - M_{jii}).
\end{equation}

\clearpage


\begin{thebibliography}{100}
\bibitem{footnote0} Our analysis can be straightforwardly extended to gauge theories with any finite Abelian gauge group $G$; we focus
on $G = (\mathbb{Z}_N)^K$ primarily to simplify the discussion.
\bibitem{footnote1} There is also another two-loop braiding process in which a small loop $\alpha$ braids around a
large loop $\beta$, in the same way that a particle braids around a loop. The statistical phase for this process is similar to
(\ref{particlestat}): $\theta_{\alpha\beta}=q_\alpha\cdot\phi_\beta$.
\bibitem{footnote3} We say that a multi-loop braiding process can be ``smoothly'' deformed into another process if we can interpolate between the two processes via a sequence of local changes to the worldsheets swept out by the loops. This sequence of local changes can be arbitrary except that the worldsheets of any two loops that are being braided around one another must stay far apart during every step.
\bibitem{footnote2} The exchange statistics $\theta_{\alpha, c}$ also satisfies linearity relations which are analogues of
(\ref{linear2}-\ref{linear3}), but we will not list them here.
\end{thebibliography}
\end{document}